\documentclass[showpacs,preprintnumbers,amsmath,amssymb]{revtex4}

\usepackage{graphicx}
\usepackage{dcolumn}
\usepackage{bm}

\begin{document}

\title{Gravitational backreaction of anti-D branes in the warped compactification}

\author{Kayoko Koyama$^{1}$ and Kazuya Koyama$^{2}$}
 
\affiliation{$^1$ Department of Phyics and Astronomy, University of Sussex, 
Brighton, BN1 9QH, UK\\
$^2$ Institute of Cosmology and Gravitation, Portsmouth 
University, Portsmouth, PO1 2EG, UK}

\date{\today}

\begin{abstract}
We derive a low-energy effective theory for gravity with anti-D branes,
which are essential to get de Sitter solutions in the type IIB string 
warped compactification, 
by taking account of gravitational backreactions of anti-D branes. 
In order to see the effects of the self-gravity of anti-D branes,  
a simplified model is studied where a 5-dimensional anti-de Sitter ({\it AdS}) spacetime 
is 
realized by the bulk cosmological constant and the 5-form flux, 
and anti-D branes are coupled to the 5-form field by Chern-Simon terms.
The {\it AdS} spacetime is truncated by introducing UV and  
IR cut-off branes like the Randall-Sundrum model.  
We derive an effective theory for gravity on the UV brane  
and reproduce the familiar result that the tensions of the anti-D branes  
give potentials suppressed by the forth-power of the warp  
factor at the location of the anti-D branes.  
However, in this simplified model, the potential energy never  
inflates the UV brane, although the anti-D-branes are inflating.   
The UV brane is dominated by dark radiation coming from the  
projection of the 5-dimensional Weyl tensor, unless the moduli fields for  
the anti-D branes are stabilized. We comment on the possibility  
of avoiding this problem in a realistic string theory compactification.  
\end{abstract}

\maketitle
\hspace{1cm}\\

\section{1.Introduction}
It has been a difficult problem to find de Sitter solutions 
in the context of string theory \cite{nogo}. 
Recently, significant progress 
has been made in the context of the type IIB string theory. 
Ref.\cite{KKLT} considered the warped compactification where 
all moduli fields are stabilized by the flux of the form fields except for the 
volume modulus for Calabi-Yau space \cite{GKP}.
The ten-dimensional metric is given by
\begin{equation}
ds_{10}=a(r)^2 g_{\mu \nu} dx^{\mu} dx^{\nu} + a(r)^{-2}
\tilde{g}_{mn} dr^m dr^n,
\end{equation}
where $a(r)$ is the warp factor. In the intermediate region, 
the warp factor is approximated as (see figure 1)
\begin{equation}
a(r)^2 = \left( \frac{r}{\ell} \right)^2.
\end{equation}
Then the spacetime looks like $AdS_5 \times X^5$ where 
$AdS_5$ is a 5-dimensional Anti-de Sitter spacetime and 
$X^5$ is a compactified manifold. For large $r$, the spacetime
deviates from $AdS_5$ spacetime and the geometry is smoothly
joined to a Calabi-Yau compactification. This gluing region plays the role
of the ultarviolet cut-off of the $AdS_5$ spacetime. The geometry
also deviates from $AdS_5$ and terminates in the far infrared although the 
spacetime remains smooth \cite{KS}. 
Then the warp factor has a minimum value at small $r=r_0$. 

\begin{figure}[h]
\centerline{
\includegraphics[width=10cm]{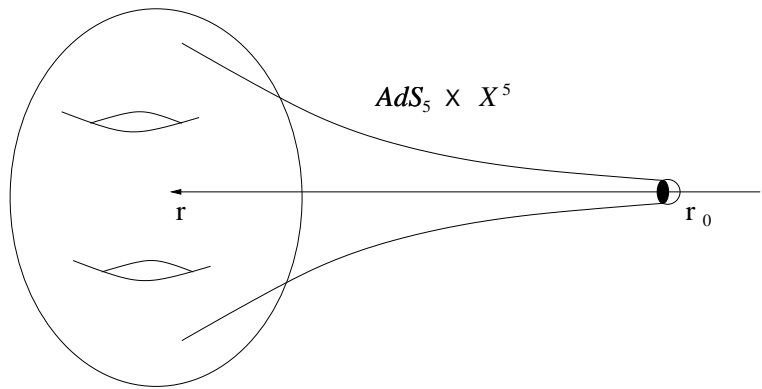}}
\caption{Schematic picture of the string warped compactification. 
The {\it AdS} spacetime is truncated at large $r$ (UV) and small $r$ (IR). 
Anti-D branes are sitting at the tip of the warped throat.}
\end{figure}

In this compactification, the vacuum solution for 4-dimensional spacetime is 
given by $AdS$ spacetime. In order to lift the spacetime to de Sitter 
spacetime, Ref.\cite{KKLT} introduces anti-D branes, which 
break the supersymmetry and give the potential energy
\begin{equation}
V_{\bar{D}}= \frac{2 a_0^4 T_3}{g_s^4} \frac{1}{\sigma^3},
\label{potential}
\end{equation}
where $a_0$ is the warp factor at the location of the anti-D 
brane, $T_3$ is the tension of the brane, $g_s$ is the string 
coupling and $\sigma$ is the volume modulus. This potential energy can 
create a de Sitter spacetime. Because the potential energy is minimized 
at small warp factor, the anti-D branes move towards $r=r_0$ and 
sit at the tip of the warped throat \cite{antiD}.  
In order to stabilize the volume modulus $\sigma$, they introduce  
non-perturbative effects. Then they get metastable de Sitter 
spacetime where all moduli-fields are stabilized.

After stabilizing the moduli fields, this model is quite similar 
to a Randall-Sundrum model, where the UV brane and the IR brane are 
introduced in $AdS_5$ spacetime by hand \cite{RS}. In this model,   
four-dimensional gravity is reproduced on the UV brane if the IR 
brane is located at the far infrared \cite{GT}. Thus 
we naively expect that we get 4-dimensional Einstein gravity
near the gluing region which plays the role of the UV cut-off brane. 

An effective theory for the anti-D brane coupled to gravity is 
discussed in Ref. \cite{KKLMMT}. They also introduced a mobile D-brane to have 
slow-roll inflation. However, their derivation is basically 
based on a probe brane approximation. A probe brane approximation is adequate 
for the calculation of the motion of a light brane in a fixed background
spacetime. However, if we want to discuss the effect of the brane 
on the gravity of the UV cut-off brane, we do need to take into 
account the gravitational back-reaction of the brane on the 
geometry. In other words, we have to take into account the 
self-gravity of the brane. Thus it would be desirable to derive
the effective gravitational theory by taking into account the 
self-gravity of the brane.

 A difficulty is that we do not have the technology to deal with the 
self-gravity of infinitely thin branes with co-dimensions higher than one.
If the co-dimension is one, we can use the junction condition to 
take into account the self-gravity \cite{Israel}. Thus in this paper, we 
use a 5-dimensional toy model proposed in Ref.\cite{Shiromizu, Shiromizu1}. 
This is a model motivated by the 5-dimensional supergravity 
of type IIB theory compactified on $AdS_5 \times S^5$ \cite{Shiromizu0}. 
In order to have $AdS_5$ spacetime as a solution, we assume 
all moduli fields, including the volume moduli for $S^5$, are stabilized. 
We also keep only the 
5-form field which is relevant for our discussions. Then 
this model is basically an extension of the Randall-Sundrum model 
with the 5-form flux. We introduce the UV and IR cut-off branes in $AdS_5$ 
spacetime and assume that standard model particles are living on the 
UV cut-off brane. This model is described by the 5D action \cite{Shiromizu}
\begin{eqnarray}
S   &=& S_5 + \sum_i ( S_i^{brane} + S_i^{CS}), \nonumber \\
S_5 &=& {1 \over 2 \kappa^2} \int d^5 x \sqrt{-{}^{(5)}g} 
       \left( {}^{(5)}R  - 2 \Lambda - {1 \over 2} |G|^2 \right), \nonumber\\
S_i^{brane}&=& -T_i \int d^4 x \sqrt{-g},   \nonumber\\
S_i^{CS}   &=& Q_i \int d^4 x \sqrt{-g} ~{1 \over 4!} ~ 
               \varepsilon^{\mu\nu\rho\sigma} D_{\mu\nu\rho\sigma}, \nonumber\\
\label{action} 
\end{eqnarray} 
where $g_{\mu \nu}$ is the induced metric on the brane, $\varepsilon^{\mu\nu\rho\sigma}$ 
is 
a four dimensional epsilon tensor for $g_{\mu \nu}$ which satisfies 
$\varepsilon^{\mu\nu\rho\sigma} \varepsilon_{\mu\nu\rho\sigma}= 
- 4!$. The 5-form field $G_{ABCDE} = {1 \over 4!} \partial_{[A} D_{BCDE]}$
has nonzero component 
$G_{y\mu\nu\rho\sigma} = \partial_{y} D_{\mu\nu\rho\sigma}$, so 
\begin{equation}
\vert G \vert^2 = {1 \over 5!} G_{ABCDE} G^{ABCDE} 
= {1 \over 4!} \partial_{y} D_{\mu\nu\rho\sigma} 
               \partial^{y} D^{\mu\nu\rho\sigma}.
\end{equation}
We impose the BPS condition on the tension and charge 
\begin{equation}
\vert T_i \vert = \vert Q_i \vert, 
\end{equation}
for each brane. We also assume $Z_2$ symmetry across the brane for the UV and IR cut-off 
branes. In order to derive the 4-dimensional effective theory on the 
UV brane, we need to solve the 5-dimensional Einstein equation. 
We use a gradient expansion method to solve the 5-dimensional 
Einstein equation \cite{grad}. 

The structure of the paper is as follows.
In section II, we introduce a D-brane in between the 
cut-off branes and study the back-reaction of the D-brane.
We show that there appears no potential if we tune the 
cosmological constant $\Lambda$ in the bulk appropriately. In section
III, we introduce an anti-D brane. In this case, the opposite 
sign of tension $T_0$ and charge $Q_0$ for the anti-D brane makes
it impossible to have a static solution. Thus we must introduce
the anti D-brane perturbatively by assuming $T_0=-Q_0$ is much smaller 
than the tension of the UV brane, $T_+$, and the tension of the IR brane, $T_-$. 
We show that the tension of the anti-D brane
gives a potential energy coupled to a moduli that represents the location
of the anti-D brane. However, we see that, unless we stabilize this 
moduli field, the UV brane never
inflates by the potential energy. Rather, the UV brane is always a radiation 
dominated universe. We provide a simple explanation for this fact 
from a geometrical point of view.
In section IV, we generalize our analysis to include 
many D and anti-D branes.  Section V is devoted to conclusions. 

\section{D brane}
\subsection{set-up}
In this section, we introduce a D-brane in the truncated $AdS_5$ 
spacetime. 
The metric is taken as 
\begin{eqnarray}
ds^2  &=& g_{yy}(x)~ dy^2 
         + g_{\mu \nu} ( x, y ) dx^{\mu} dx^{\nu} \nonumber \\
     &=& e^{2 \phi(x)} dy^2 
         + e^{-2 u(x,y)} h_{\mu \nu} ( x, y ) dx^{\mu} dx^{\nu},
\label{metric} 
\end{eqnarray} 
where $\phi(x)$ is an arbitrary function of $x$ and 
$g_{\mu \nu}$ is the induced metric of 
$y$= constant hypersurfaces. The UV brane, D3 brane and IR brane
are located at $y = y_{+}, y_0$ and $y_{-}$ respectively. 
The warp factor $u(x, y)$ is an arbitrary function of $x$ and $y$ at this 
stage.  The physical distances between two branes are given by 
$$
d_{+}(x) = \int_{y_+}^{y_0} dy ~e^{\phi}, ~~~~~~~~
d_{-}(x) = \int_{y_0}^{y_{-}} dy ~e^{\phi} ,
$$
which are called radions. We use the index 
$i=(+,0,-)$ to denote the variables related to the branes
at $y_{+}, y_0$ and $y_{-}$ respectively. 
We assume all three branes satisfy
\begin{equation}
T_i = Q_i.
\end{equation} 
We consider the bulk with 
curvature radius and cosmological constant given by
$l_+ , \Lambda_+ $ respectively for $y=(y_+,y_0)$ and 
$l_- , \Lambda_- $ for $y=(y_0,y_-)$ (see figure 2).

\begin{figure}[h]
\centerline{
\includegraphics[width=10cm]{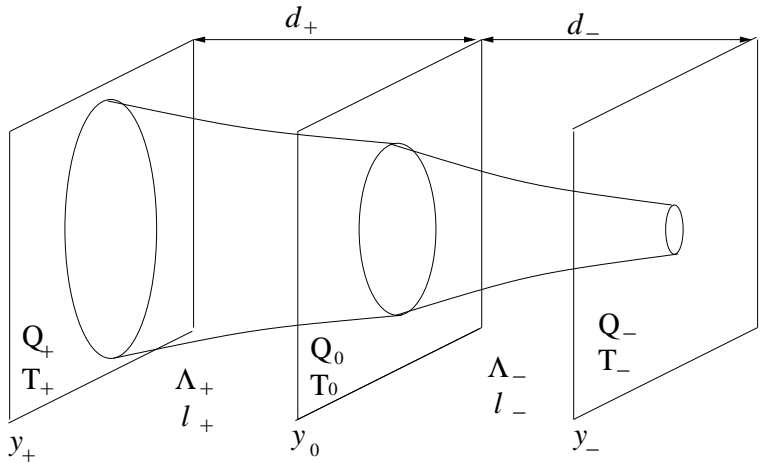}}
\caption{The case of a D-brane in the middle of the cut-off branes.
The tension of the D-brane is positive for $l_+ > l_-$. }
\end{figure} 

\subsection{Field equations and perturbative solutions}
Firstly, the field equation for the form field is given by
\begin{eqnarray}
\partial_y ( \sqrt{-{}^{(5)}g} ~\partial^y D^{\mu\nu\rho\sigma} ) 
= -2 \kappa^2 \sqrt{-g} ~\varepsilon^{\mu\nu\rho\sigma}(g)
\sum_i Q_i \delta(y-y_i),
\end{eqnarray}
where $\epsilon_{\mu \nu \rho \sigma}(g)$ is a epsilon tensor 
defined with respect to $g_{\mu \nu}(y,x)$.
The junction conditions at $ y_+$ and $y _- $ are given by
\begin{eqnarray}
\partial_y D^{\mu\nu\rho\sigma}
= \left\{
\begin{array}{l}
- \kappa^2 \varepsilon^{\mu\nu\rho\sigma}(g(y_+,x)) Q_+ \vspace{0.3cm}, \\
+ \kappa^2 \varepsilon^{\mu\nu\rho\sigma}(g(y_-,x)) Q_- ,
\end{array} \right.
\end{eqnarray}
We can easily find the solutions that satisfy these boundary conditions as
\begin{eqnarray}
\partial_y D^{\mu\nu\rho\sigma}
= \left\{
\begin{array}{l}
- \kappa^2 e^{4 u(y,x)}\varepsilon^{\mu\nu\rho\sigma}(h) Q_+, \:\: y_+ <y <y_0, 
\vspace{0.3cm} \\
+ \kappa^2 e^{4 u(y,x)}\varepsilon^{\mu\nu\rho\sigma}(h) Q_-, \:\: y_0 < y < y_-.
\end{array} \right.,
\end{eqnarray}
where $\varepsilon_{\mu \nu \rho \sigma}(h)$ is a epsilon tensor defined 
with respect to $h_{\mu \nu}$. 
The junction condition at $y_0 $ is given by
\begin{equation}
-\partial_y D^{\mu\nu\rho\sigma}|_{y_0 + \delta} + 
\partial_y D^{\mu\nu\rho\sigma}|_{y_0 - \delta} = 2 \kappa^2 
\varepsilon^{\mu\nu\rho\sigma}(g(y_0,x)) Q_0,
\end{equation}
where $\delta \to 0$. 
This gives a constraint on the charges of three branes, 
\begin{eqnarray}
Q_+ + Q_- + 2 ~Q_0 = 0.   
\end{eqnarray}

Next, we consider the Einstein equations. 
We define the extrinsic curvature on the $y$=constant slicing as
\begin{eqnarray}
K_{\mu\nu} = \displaystyle{\frac{1}{2} e^{-\phi} \partial_y g_{\mu\nu} }= 
\left\{
\begin{array}{ll}
\displaystyle{K^+_{\mu\nu}}, ~~~
& y_+ < y <y_0, \vspace{0.3cm} \\
\displaystyle{K^-_{\mu\nu}}, ~~~
& y_0 < y< y_-, 
\end{array} \right.
\end{eqnarray}
to solve those Einstein equations by the gradient expansion method.
It is useful to decompose the extrinsic curvature 
into a trace part 
$K= K^{\mu}_{\nu}$ 
and
a traceless part 
$ \widetilde{K}^{\mu}_{\nu} = K^{\mu}_{\nu} -{1 \over 4} g_{\mu\nu} K $.
Then the Einstein equations can be written as
\begin{eqnarray}
&&\partial_{y} \widetilde{K}^{\mu}_{\nu} 
=  - K \widetilde{K}^{\mu}_{\nu}  
+  
   {}^{(4)}\widetilde{R}^{\mu}_{\nu} -\kappa^2 \left(  
{}^{(5)} T^{\mu}_{\nu} - {1 \over 4} \delta ^{\mu}_{\nu} {}^{(5)} 
T^{\rho}_{\rho} \right)
 + \Phi^{\mu}_{\nu} , \\
&&\partial_{y} K =  - K^2 
+   {}^{(4)}R 
                    + {\kappa^2 \over 3} ({}^{(5)} T^{\mu}_{\mu} + 4 
					{}^{(5)} T^y_y)
                    -\nabla^2 \phi - ( \nabla \phi )^2 , \\
&& \widetilde{K}_{\mu\nu} \widetilde{K}^{\mu\nu}
- {3 \over 4} K^2 +  {}^{(4)}R = - 2 \kappa^2 {}^{(5)} T_{yy}, 
\label{Hamiltonian} \\
&& \nabla_{\nu} K^{\nu}_{\mu} + \nabla_{\mu} K = 0, 
\end{eqnarray}
where
$$
\Phi^{\mu}_{\nu} = - \nabla^{\mu} \nabla_{\nu} \phi  
 + {1 \over 4} \delta ^{\mu}_{\nu} \nabla^2 \phi
 -  (\nabla^{\mu} \phi) (\nabla_{\nu} \phi)  
  + {1 \over 4} \delta ^{\mu}_{\nu} ( \nabla \phi )^2,
$$
$\nabla_{\mu}$ denotes the covariant derivative 
with respect to the metric $h_{\mu\nu}$,  
$ {}^{(4)} R^{\mu}_{\nu} $ is the corresponding 4-dimensional Ricci 
curvature and ${}^{(5)} T_{AB}$ is the 5-dimensional energy-momentum tensor. 
We also decompose the 4-dimensional curvature into a trace part $R = R^{\mu}_{\nu}$ and
a traceless part 
$ \widetilde{R}^{\mu}_{\nu} = R^{\mu}_{\nu} -{1 \over 4} g_{\mu\nu} R $.
The junction conditions are given by
\begin{eqnarray}
&& \left( K_{\mu \nu} - K g_{\mu \nu}\right) |_{y_\pm}=\pm \frac{\kappa^2}{2} 
T_{\pm}g_{\mu \nu},\\
&& -(K_{\mu \nu} - K g_{\mu \nu}) |_{y_0 + \delta} +(K_{\mu \nu} - K g_{\mu \nu})
|_{y_0 - \delta} = -\kappa^2 T_0 g_{\mu \nu},
\end{eqnarray}
where $\delta \to 0$.

We derive a low energy theory for this three-brane system
by the gradient expansion method \cite{grad}. We assume that the 4-dimensional curvature 
and the derivative of $\phi(x)$ with respect to brane coordinates 
are suppressed as 
\begin{equation}
{}^{(4)}R \:\: l_+^2 \sim (\nabla_{\mu} \phi) \:\: l_+ \sim \epsilon \ll 1.  
\end{equation}
Then the extrinsic curvature can be expanded in terms of $\epsilon \ll 1$
\begin{equation}
K^{\mu}_{\nu}= {}^{(0)} K^{\mu}_{\nu}+{}^{(1)} K^{\mu}_{\nu}+ \cdot \cdot .
\end{equation}
We can solve the Einstein equations perturbatively in $\epsilon$.

\hspace{1cm}\\
(1) Zeroth order\\

At 0-th order, we can put the warp factor in Eq. (\ref{metric}) 
in the following form 
\begin{eqnarray}
u(x,y) = 
\left\{
\begin{array}{ll}
\displaystyle{\frac{e^{\phi} ~y}{l_+}},  & y_+ < y <y_0, \vspace{0.3cm} \\
\displaystyle{\frac{e^{\phi} ~y}{l_-} 
   - e^{\phi} ~y_0 \left({1 \over l_-} - {1 \over l_+}\right)}, & y_0 < y< y_-.
\end{array} \right.
\end{eqnarray} 
The 0-th order solutions for $K_{\mu \nu}$ are obtained as 
\begin{eqnarray}
&&{}^{(0)} K_{\pm} = -{4 \over l_{\pm}}, \\
&&{}^{(0)} {K_{\pm}}^{\mu}_{\nu} = -{1 \over l_{\pm} } \delta^{\mu}_{\nu}.
\end{eqnarray}
Substituting these solutions into the "Hamiltonian constraint" (\ref{Hamiltonian}), 
we get 
\begin{equation}
- {6 \over l^2_{\pm}} = \Lambda_{\pm} + {1 \over 4}\kappa^4 ~Q_{\pm}^2.  
\end{equation}
We can think of this as conditions on $\Lambda_{\pm}$. 
On the other hand, the junction conditions give 
\begin{eqnarray}
&&{ 6 \over l_{\pm}} = \pm \kappa^2 T_{\pm}, \\
&& T_+ +T_- + 2~ T_0 = 0.
\end{eqnarray}
We should note that the constraints for the tensions are consistent with 
the conditions for the charges if all three branes are D-branes, 
$Q_i = T_i$. 

\hspace{1cm}\\
(2) First order\\

At first order, we obtain the solutions as 
\begin{eqnarray}
&&{}^{(1)} K= -{l_{\pm} \over 6}  R, \\
&&{}^{(1)} \widetilde{K}^{\mu}_{\nu} = 
- {l_{\pm} \over 2}  \widetilde{R}^{\mu}_{\nu}
- e^{4 e^{\phi} y/l_{\pm}}  \chi^{\mu}_{\nu}(l_{\pm},x),
\end{eqnarray}
where $R$ and $R^{\mu}_{\nu}$ are defined with respect to the 
metric $g_{\mu \nu}(y,x)$. 
The integration constants
\begin{eqnarray}
\chi_{\mu\nu} (x) =
\left\{
\begin{array}{ll}
\chi_{\mu\nu}(l_+,x), ~~~& y_+ < y <y_0, \vspace{0.3cm} \\
\chi_{\mu\nu}(l_-,x), ~~~& y_0 < y< y_-, 
\end{array} \right.
\end{eqnarray}
do not depend on $y$ and satisfy
$$
\chi^{\mu}_{\mu}(l_{\pm},x) = 0 ~~~,~~~
\nabla_{\mu} \chi^{\mu}_{\nu}(l_{\pm},x) = 0. 
$$
The junction conditions at $y_{\pm}$ and $y_0$ give the equations
\begin{eqnarray}
&&{l_{\pm} \over 2} G^{\mu}_{\nu}( l_{\pm},y_{\pm})  
+ e^{4 e^{\phi} y_{\pm}/l_{\pm}} \chi^{\mu}_{\nu} (l_{\pm},x) = 0, 
\label{eq1}
\\
&&\Big[{l_+ \over 2} G^{\mu}_{\nu}( l_+,y_0)  
+ e^{4 e^{\phi} y_0 /l_+} \chi^{\mu}_{\nu}(l_+,x)\Big]
- \Big[ {l_- \over 2} G^{\mu}_{\nu}( l_-,y_0)  
+ e^{4 e^{\phi} y_0 /l_-} \chi^{\mu}_{\nu}(l_-,x) \Big]= 0.
\label{eq2}
\end{eqnarray}

\subsection{Effective equations on the UV brane}
Now we are ready to derive the effective equations on the UV brane 
at $y=y_+$.
One can eliminate the integration constants from Eqs.~(\ref{eq1},\ref{eq2})
to find the following equation
\begin{eqnarray}
  {l_+ \over 2} \left[ G^{\mu}_{\nu}( l_+,y_0)  
- e^{4 d_{+}/l_{+}} G^{\mu}_{\nu}(l_+,y_+) \right]
- {l_- \over 2} \left[ G^{\mu}_{\nu}(l_-,y_0)  
- e^{-4 d_-/l_-} G^{\mu}_{\nu}(l_-,y_-) \right] = 0,
\end{eqnarray}
where $G^{\mu}_{\nu} (l,y)$ denotes the 4-dimensional Einstein tensor 
on a $y$= constant hypersurface of the bulk with curvature length $l$.  
We can rewrite this equation in terms of the induced metric of the UV brane
at $y_+$. The Einstein tensor satisfies the following relations:
\begin{eqnarray}
G^{\mu}_{\nu}( l_+,y_+) & \equiv & {}^{(4)}G^{\mu}_{\nu},  \\
G^{\mu}_{\nu}( l_+,y_0) &=& 
G^{\mu}_{\nu}( l_-,y_0) 
 =  e^{2 d_+/l_+} \Big[ {}^{(4)}G^{\mu}_{\nu} 
  + {\cal F}^{\mu}_{\nu} ( l_+, d_+) \Big],    \\
G^{\mu}_{\nu}( l_-,y_-) 
 &=&  e^{ - 2 d_+/l_+ - 2 d_-/l_-} \Big[ {}^{(4)}G^{\mu}_{\nu} 
  + {\cal F}^{\mu}_{\nu} ( l_+, d_+) 
  + {\cal F}^{\mu}_{\nu} ( l_-, d_-) 
  + f^{\mu}_{\nu} (l_+,d_+; l_-, d_-) \Big],
\end{eqnarray}
where 
\begin{eqnarray}
{\cal F}^{\mu}_{\nu}(l,d) &=& 
\frac{2}{l} \left( \nabla^{\mu} \nabla_{\nu} d - \delta^{\mu}_{\nu} \nabla^2 d \right)
+\frac{2}{l^2} \left(\nabla^{\mu}d \nabla_{\nu} d+ 
\frac{1}{2} \delta^{\mu}_{\nu} (\nabla d)^2 \right), \\
f^{\mu}_{\nu}(l_+,d_+; l_-, d_-) &=& 
{2 \over {l_+ l_-}} ( 
\nabla^{\mu} d_+ \nabla_{\nu} d_-
+ \nabla^{\mu} d_- \nabla_{\nu} d_+ 
+ \delta^{\mu}_{\nu} \nabla_{\rho} d_+ \nabla^{\rho} d_-).
\end{eqnarray}
Then the effective equations on the UV brane are obtained as 
\begin{eqnarray}
\Big[1 &-& C_0 e^{-2 d_+/l_+} - C_- e^{- 2 (d_+/l_+ +d_-/l_-)} \Big]
{}^{(4)}G^{\mu}_{\nu} \nonumber\\
&=& C_0 e^{-2 d_+/l_+} {\cal F}^{\mu}_{\nu} ( l_+, d_+) 
+ C_- e^{-2 (d_+/l_+ + d_-/l_-)} \Big[ 
{\cal F}^{\mu}_{\nu} ( l_+,  d_+) + {\cal F}^{\mu}_{\nu} ( l_-,  d_-)
+f^{\mu}_{\nu}(l_+, d_+; l_-, d_-) \Big], 
\end{eqnarray}
where 
\begin{equation}
C_0 = \frac{l_+-l_-}{l_+}, \quad C_-=\frac{l_-}{l_+}.
\end{equation}
The field equations for the radions are obtained from the traceless 
conditions for $\chi^{\mu}_{\nu}(l_{\pm},x)$ as 
\begin{eqnarray}
&& {\cal F}^{\mu}_{\mu} ( l_+, d_+) = 0, \\
&& {\cal F}^{\mu}_{\mu} ( l_+,  d_+) + {\cal F}^{\mu}_{\mu} ( l_-,  d_-)
+f^{\mu}_{\mu}(l_+, d_+; l_-, d_-) = 0.
\end{eqnarray}

These effective equations can be derived from the following 
effective action
\begin{equation}
S_{eff} = \frac{1}{2 \kappa_4^2} \int d^4x \sqrt{-h} \Big[ 
(1- C_0 \Phi_0^2 - C_- \Phi_-^2 ) {}^{(4)}R 
- 6 C_0 \partial_{\mu} \Phi_0 \partial^{\mu} \Phi_0 
- 6 C_- \partial_{\mu} \Phi_- \partial^{\mu} \Phi_- \Big],
\end{equation}
where $\kappa_4^2 =\kappa^2/l_+ $ and we defined 
two scalar fields as
\begin{equation}
\Phi_0 \equiv e^{-d_+/l_+}, \quad 
\Phi_- \equiv e^{-d_+/l_+-d_-/l_-}.
\end{equation}
The over-all normalization was determined by substituting the 
solutions into the 5-dimensional action. This effective action is 
the same as the one for a three-brane system in the Randall-Sundrum 
model \cite{LW}.  
If we take the limit $l_+ -l_- \to 0$, $C_0=0$ and the middle D-brane 
disappears. Then we reproduce the familiar two brane result \cite{grad}.

We should mention that the tension of the middle brane is 
given by
\begin{equation}
T_0=-\frac{1}{2}(T_+ + T_-) = -\frac{3}{\kappa^2}
\left(\frac{1}{l_+} - \frac{1}{l_-}\right).
\end{equation}
Thus if the middle brane has a positive tension
we should have $l_+ > l_-$. In this case $C_0 >0$ and kinetic 
terms for $\Phi_0$ and $\Phi_-$ are normal. If the middle 
brane has a negative tension, $C_0 < 0$. In order to 
see the kinetic term for scalar fields, it is better 
to move to the Einstein frame. By defining new fields and 
metric \cite{LW},
\begin{eqnarray}
\chi &=& - \ln \left\vert \frac{1- \sqrt{1-\varphi}}{1+ \sqrt{1 - \varphi}}   
\right\vert,
\quad \varphi = 1- C_0^2 \Phi_0^2 - C_-^2 \Phi_-^2,\\
\psi &=& 
\left\{
\begin{array}{ll}
\displaystyle{\ln \left \vert  \frac{z-1}{z+1}  \right \vert, \quad z = 
\sqrt{-\frac{C_0}{C_-}}
\frac{\Phi_0}{\Phi_-}}, ~~~
& C_0 <0, \vspace{0.3cm} \\
\displaystyle{2 \arctan z, \quad z =  \sqrt{\frac{C_0}{C_-}}
\frac{\Phi_0}{\Phi_-}},~~~
& C_0 >0,
\end{array} \right.\\
g_{E\:\mu\nu} &=& \varphi h_{\mu \nu},
\end{eqnarray}
the effective action can be rewritten as 
\begin{equation}
S = \frac{1}{2 \kappa_4^2} \int d^4 x \sqrt{-g_E} 
\left[ R_E - \frac{3}{2} \partial_{\mu} \chi \partial^{\mu} \chi 
\pm \frac{3}{2}  \left( \sinh^2 \frac{\chi}{2} \right)  
\partial_{\mu} \psi \partial^{\mu} \psi 
\right],
\end{equation}
where $+$ for $C_0 <0$ and $-$ for $C_0 >0$. If we consider fluctuations 
from static solutions $R=R_E=0, \psi=\chi=$ const., the fluctuation for 
the middle D-brane becomes a ghost if $C_0 <0$, i.e. if the middle brane has a 
negative tension \cite{radion}. 
We assume the middle D-brane has a positive tension. 

Because there are two independent conformal coupled scalar fields,
if we consider a homogeneous and isotropic universe, the Friedmann equation on the UV 
brane is given by
\begin{equation}
H^2 = C a_4^{-4},
\end{equation}
where $C$ is an integration constant and $a_4$ is the scale factor 
of the universe on the UV brane. Thus the universe is dominated by 
the radiation, which is called dark radiation. The integration 
constant $C$ is related to the Black Hole mass in the bulk. 
This is consistent with the effective equations given by
\begin{equation}
{l_{+} \over 2} {}^{(4)} G^{\mu}_{\nu} 
=-\chi^{\mu}_{\nu} (l_+,x),
\end{equation}
where $\chi^{\mu}_{\nu}$ is a transverse-traceless tensor.  
We can relate $\chi^{\mu}_{\nu}$ to the electric part of the 5-dimensional 
Weyl tensor projected on the brane, $E^{\mu}_{\nu}$, as \cite{SMS}
\begin{equation}
\chi^{\mu}_{\nu} (l_+,x)=  {l_{+} \over 2} E^{\mu}_{\nu}, \quad
E_{\mu \nu} = {}^{(5)} C^{y}_{\:\: \mu y \nu}.
\end{equation}

\section{Anti-D brane}
Now, we introduce an anti-D brane 
at $y=y_0$ instead of a D brane. The relation between the tension
and the charge is given by
\begin{equation}
T_0=-Q_0,
\end{equation}
for the anti-D brane. 
We can immediately see that there is no static solution. 
The constraint on the charges is unchanged and this can be rewritten 
in terms of the tensions as
\begin{equation}
Q_+ + Q_- +2 Q_0 =T_+ + T_- - 2 T_0=0,
\label{chargecon}
\end{equation}
where we assumed $T_{\pm}=Q_{\pm}$. This contradicts the condition
to have a static solution, given by $T_+ + T_- + 2 T_0=0$.
Thus we assume $T_0 \ll T_{\pm}$ and treat $T_+ + T_- + 2 T_0$ as a first order 
quantity in the gradient expansion.

We arrive at the same equations as the D brane case but the 
junction condition at the anti-D brane is modified as 
\begin{equation}
\Big[{l_+ \over 2} G^{\mu}_{\nu}( l_+,y_0)  
+ e^{4 e^{\phi} y_0 /l_+} \chi^{\mu}_{\nu}(l_+,x)\Big]
- \Big[{l_- \over 2} G^{\mu}_{\nu}( l_-,y_0)  
+ e^{4 e^{\phi} y_0 /l_-} \chi^{\mu}_{\nu}(l_-,x) \Big]= \frac{1}{2} \kappa^2 \Gamma_0 
\delta^{\mu}_{\nu} ,
\label{eqanD}
\end{equation}
where 
\begin{equation}
\Gamma_0 = T_+ + T_- + 2 T_0 = 4 T_0.
\end{equation}
Here we used Eq.~(\ref{chargecon}). Because the tension of the anti-D brane 
is assumed to be small compared with $T_{\pm}$, we need to assume 
$\vert l_+ - l_- \vert \ll l_{\pm}$. 
Then we can get the effective equations on the 
positive tension branes as 
\begin{eqnarray}
\Big[1 &-& C_0 e^{-2 d_+/l_+} - C_- e^{- 2 (d_+/l_+ +d_-/l_-)} \Big]
{}^{(4)}G^{\mu}_{\nu} \nonumber\\
&=& C_0 e^{-2 d_+/l_+} {\cal F}^{\mu}_{\nu} ( l_+, d_+)
+ C_- e^{-2 (d_+/l_+ + d_-/l_-)} \Big[ 
{\cal F}^{\mu}_{\nu} ( l_+,  d_+) + {\cal F}^{\mu}_{\nu} ( l_-,  d_-)
+f^{\mu}_{\nu}(l_+, d_+; l_-, d_-) \Big] \nonumber\\
&& - 4 T_0 \kappa_4^2  e^{-4 d_+/l_+} \delta^{\mu}_{\nu}.
\end{eqnarray}
The equations for the radions are given by
\begin{eqnarray}
&& (l_+-l_-)
{\cal F}^{\mu}_{\mu} ( l_+, d_+) = 16 \kappa^2 T_0
e^{-2 d_+/l_+}, 
\label{eqrad}
\\
&& {\cal F}^{\mu}_{\mu} ( l_+,  d_+) + {\cal F}^{\mu}_{\mu} ( l_-,  d_-)
+f^{\mu}_{\mu}(l_+, d_+; l_-, d_-)= 0.
\end{eqnarray}

The effective action is given by
\begin{equation}
S_{eff} = \frac{1}{2 \kappa_4^2} \int d^4x \sqrt{-h} \Big[ 
(1- C_0 \Phi_0^2 - C_- \Phi_-^2 ) {}^{(4)}R 
- 6 C_0 \partial_{\mu} \Phi_0 \partial^{\mu} \Phi_0 
- 6 C_- \partial_{\mu} \Phi_- \partial^{\mu} \Phi_- 
-8 \kappa_4^2 T_0 \Phi_0^4 \Big].
\end{equation}

Then we can identify the effective cosmological constant on the UV brane 
as 
\begin{equation}
\Lambda_4 = 4 \kappa_4 T_0 \Phi_0^4 = 4 \kappa_4 T_0 e^{-4 d_+/l_+}.
\end{equation}
This is the desired result. The effective cosmological constant 
on the UV brane is proportional to the tension of the 
anti-D brane and it is suppressed by the fourth power of the 
warp factor at the location of the anti-D brane. 

The tension of the anti-D brane is given by
\begin{equation}
T_0 = \frac{1}{2} (T_+ + T_-) =\frac{3}{\kappa^2}
\left(\frac{1}{l_+} - \frac{1}{l_-} \right).
\end{equation}
Thus, for the anti-D brane, we need  $l_+ < l_-$ 
to have a positive tension brane and a positive cosmological constant (see 
Figure 3). 
Then we have $C_0 <0$ and, in the Einstein frame, the kinetic term for $\psi$ 
is always positive. Thus there would be a ghost-like excitation, although the 
situation is more complicated than the D-brane case because the tension of 
the anti-D brane gives non-trivial potentials for $\chi$ and $\psi$ given by
\begin{equation}
V(\chi,\psi)=-8 \kappa_4^2 T_0 C_0^{-2} \left( \sinh^4 \frac{\chi}{2} \right) 
\left( \sinh^4 \frac{\psi}{2} \right).
\end{equation}
Then there is no static solution. At least classically, the positivity of the 
kinetic term does not lead to instability of the system.

\begin{figure}[h]
\centerline{
\includegraphics[width=10cm]{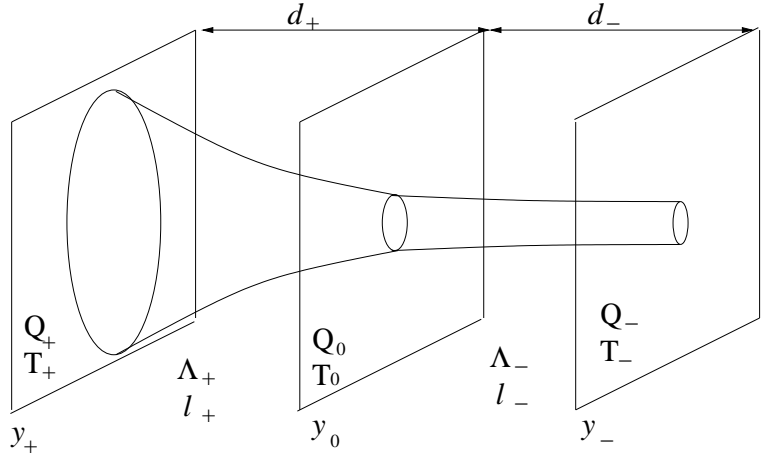}}
\caption{The case of an anti-D-brane in the middle of the cut-off branes.
The tension of the anti-D-brane is positive for $l_+ < l_-$. }
\end{figure} 

On the anti-D brane, a de Sitter spacetime is realized because 
the trace of Eq. (\ref{eqanD}) gives
\begin{equation}
R(y_0,l_+) = 16 \frac{\kappa^2}{l_--l_+} T_0,
\end{equation}  
where we assume $T_0 > 0$ and $l_+ < l_-$. 
However, unfortunately, this effective cosmological constant never inflates the 
UV brane. The Ricci scalar of the UV brane is given by
\begin{equation}
{}^{(4)} R = e^{-2 d_+/l_+} R(l_+, y_0) + {\cal F}^{\mu}_{\mu}(l_+,d_+).
\end{equation}
Using Eq. (\ref{eqrad}), we can show ${}^{(4)} R =0$. 
This fact can be easily understood by 
noting that the effective theory on the UV brane is still 
given by
\begin{equation}
{l_{+} \over 2} {}^{(4)} G^{\mu}_{\nu} 
=-\chi^{\mu}_{\nu} (l_+,x),
\end{equation}
where $\chi^{\mu}_{\nu}$ is a transverse-traceless tensor.
Thus we cannot have a non-zero Ricci scalar.
The UV brane is a radiation dominated universe 
if the brane is homogeneous and isotropic.  
We can also understand this result by the fact that the moduli 
describing the location of the anti-D brane is a conformal coupled 
scalar field and the potential preserves the conformal 
symmetry. This is related to the difficulty to get 
slow-roll inflation from an interaction between D-brane and anti-D branes \cite{KKLMMT}. 

The geometrical approach here gives a simple explanation for this. 
In {\it AdS} spacetime with an empty bulk, the radion can induce 
only dark radiation on the brane 
if we consider a homogeneous and isotropic brane. In order to 
get a de Sitter spacetime, we must introduce some matter in 
the bulk to give a potential for $d_+$. Once $d_+$ is stabilized
by bulk matter fields, we can have inflation on the UV brane 
supported by the tension of the anti-D brane. 
In a realistic string theory model, the spacetime deviates 
from {\it AdS} in the far infrared and the {\it AdS} region terminates smoothly. 
Then the anti-D brane will sit at the tip of the warped throat.
This could effectively stabilize the $d_+$ moduli. 

\section{Many branes}
Our result can be easily extended to a many-brane system \cite{Many}. 
The constraint on charges is given by
\begin{equation}
Q_+ + 2 \sum_{i=0}^{n-1} Q_i + Q_- = 0,
\end{equation}
where we have $n$ branes in between the cut-off branes. 
This can be rewritten as a condition on the tensions,
\begin{equation}
T_+ + 2 \sum_{i=D} T_i - 2 \sum_{j=\bar{D}} T_j + T_- =0,
\label{conT}
\end{equation}
where $i=D$ denotes that the branes are D branes $T_i = Q_i$
and $j=\bar{D}$ denotes that the branes are anti-D branes,  
$T_i = - Q_i$. We defined the tensions in terms of the curvature
scales $l_i$ as 
\begin{eqnarray}
T_i = 
\left\{
\begin{array}{ll}
\displaystyle{-\frac{3}{\kappa^2} \left( \frac{1}{l_{i-1}} - \frac{1}{l_i} \right)} ~~~
& \mbox{for} \:\: D \vspace{0.3cm} \\
\displaystyle{\frac{3}{\kappa^2} \left( \frac{1}{l_{i-1}} - \frac{1}{l_i} \right)} ~~~
& \mbox{for} \:\: \bar{D} 
\end{array} \right.,
\end{eqnarray}
where $l_{0}=l_+$, $l_{n}=l_-$ 
and assume that all branes have positive tensions. 
These definitions satisfy the constraint Eq. (\ref{conT}). 
We also define the cosmological constant $\Lambda_i$ in the bulk by 
\begin{equation}
- {6 \over l^2_{i}} = \Lambda_{i} + {1 \over 4}\kappa^4 ~Q_{i}^2.  
\end{equation}
Then we can easily derive the effective action on the UV branes as
\begin{equation}
S= \frac{1}{2 \kappa_4^2} \int dx^4 \sqrt{-{}^(4) g} 
\left [ \left(1-\sum_{i=0}^{n} C_i \Phi_i^2 \right) R - 6 \sum_{i=0}^{n} C_i 
\partial_{\mu} \Phi_i \partial^{\mu} \Phi_i - 8 \kappa_4^2 
\sum_{j=\bar{D}} T_j \Phi_j^4 \right],
\end{equation}
where 
\begin{equation}
\Phi_i = \exp \left(-\sum_{j=1}^{i} d_{j-1}/l_{j-1} \right), \quad
C_i = \frac{l_{i}-l_{i+1}}{l_+},
\end{equation}
$l_{n+1}=0$. $d_i$ is the physical distance between the $(i-1)$-th 
and $i$-th branes for $0 \leq i \leq n$, where we define the UV brane 
and IR brane as the $-1$-th brane and $n$-th brane respectively. 
The contribution for the potential is a linear
summation of contributions only from anti-D branes.  

 \section{Conclusion}
In this paper, we studied a low energy effective theory 
for gravity in the warped compactification model with anti-D branes.
We constructed a simplified model where the UV (positive tension) and 
the IR (negative tension) cut-off branes are introduced as in the Randall-Sundrum 
model and the standard model particles are assumed to live on the 
UV brane. Then we introduce anti-D brane(s) into the truncated 
$AdS$ spacetime. Due to the opposite sign of the tension and charge of the 
anti-D brane, the tension of the anti-D brane provides a potential 
energy suppressed by $a_0^{4}$, where $a_0$ is the warp factor
at the location of the anti-D brane. 
This could act as a cosmological constant. However, because 
the moduli field describing the distance between the UV brane and the anti-D brane is
a conformal coupled scalar field and the potential preserves the conformal
symmetry, this potential energy does not inflate the UV brane. Rather,
the UV brane is a radiation dominated universe. From a geometric point of view,
the UV brane is dominated by the energy density of the projected 5-dimensional 
Weyl tensor, known as dark radiation. 

We should note that the anti-D brane itself is inflating. 
Thus, the extra-dimension is quite inhomogeneous. Precisely 
speaking, the inhomogeneity originates from the inhomogeneous 
choice of the slicings of the $AdS$ spacetime. 
The 4-dimensional slicings of $AdS_5$ can have different geometries 
on them. Without branes, different choices of the slicing are merely 
different choices of the coordinates. However, once we introduce self-gravitating 
branes, the differences of the slicing become physical in the sense that 
the geometries of the branes are determined by the slicings. 
This is a crucial difference from the conventional Kaluza-Klein compactification
where the 4-dimensional spacetime has the same geometry regardless of 
position in the extra-dimension. 

In order to have a de Sitter solution for the UV brane, we must 
stabilize the moduli fields for anti-D branes. 
In a realistic string theory compactification, the $AdS$ spacetime terminates 
smoothly in the far infrared. 
This can act as a stablization mechanism for anti-D branes because the 
anti-D branes will sit at the tip of the throat where the warp factor is 
minimized. In order to address this
issue more precisely, we need to take into account the geometry of 
the tip instead of introducing the IR cut-off brane by hand \cite{antiD,antiD1}. 
This deserves further investigation.

\acknowledgments
We would like to thank D. Wands for informing us of Ref.\cite{LW}, 
R. Maartens for a careful reading of this manuscript and 
T. Shiromizu for pointing out several typos. 
KK is supported by PPARC. 

\vspace{1cm}


\begin{references}
\bibitem{nogo}
J.~M.~Maldacena and C.~Nunez,
Int.\ J.\ Mod.\ Phys.\ A {\bf 16} (2001) 822.

\bibitem{KKLT}
S.~Kachru, R.~Kallosh, A.~Linde and S.~P.~Trivedi,
Phys.\ Rev.\ D {\bf 68} (2003) 046005.

\bibitem{GKP}
S.~B.~Giddings, S.~Kachru and J.~Polchinski,
Phys.\ Rev.\ D {\bf 66} (2002) 106006.

\bibitem{KS}
I.~R.~Klebanov and M.~J.~Strassler,
JHEP {\bf 0008}, (2000) 052.

\bibitem{antiD}
S.~Kachru, J.~Pearson and H.~Verlinde,
JHEP {\bf 0206} (2002) 021.

\bibitem{RS}
L. Randall and R. Sundrum, Phys. Rev. Lett. {\bf 83}, (1999) 3370;
ibid, {\bf 83}, (1999) 4690. 

\bibitem{GT}
J.~Garriga and T.~Tanaka,
Phys.\ Rev.\ Lett.\  {\bf 84} (2000) 2778,

\bibitem{KKLMMT}
S.~Kachru, R.~Kallosh, A.~Linde, J.~Maldacena, L.~McAllister and S.~P.~Trivedi,
JCAP {\bf 0310} (2003) 013.

\bibitem{Israel}
W. Israel,
Nuovo Cim. B {\bf 44S10} (1966) 1.

\bibitem{Shiromizu}
T.~Shiromizu, K.~Koyama and T.~Torii,
Phys.\ Rev.\ D {\bf 68} (2003) 103513;
Y.~Iwashita, T.~Shiromizu, K.~Takahashi and S.~Fujii,
Phys.\ Rev.\ D {\bf 71} (2005) 083518.

\bibitem{Shiromizu1}
S.~Onda, T.~Shiromizu, K.~Koyama and S.~Hayakawa,
Phys.\ Rev.\ D {\bf 69} (2004) 123503;
T.~Shiromizu, Y.~Himemoto and K.~Takahashi,
Phys.\ Rev.\ D {\bf 70} (2004) 107303;
T.~Shiromizu, K.~Takahashi, Y.~Himemoto and S.~Yamamoto,
Phys.\ Rev.\ D {\bf 70} (2004) 123524;
K.~Takahashi and T.~Shiromizu,
Phys.\ Rev.\ D {\bf 70} (2004) 103507.

\bibitem{Shiromizu0}
M.~Sato and A.~Tsuchiya,
Prog.\ Theor.\ Phys.\  {\bf 109} (2003) 687;
T.~Shiromizu, K.~Koyama, S.~Onda and T.~Torii,
Phys.\ Rev.\ D {\bf 68} (2003) 063506.

\bibitem{grad}
T. Wiseman, Class. Quantum Grav. {\bf 19}, (2002) 3083;
S. Kanno and J. Soda, Phys. Rev.\ D {\bf 66}, (2002) 043526;
ibid {\bf D66}, (2002) 083506; 
T. Shiromizu and K. Koyama, Phys. Rev. {\bf D67}, (2003) 084022.

\bibitem{radion}
I.~I.~Kogan, S.~Mouslopoulos, A.~Papazoglou, G.~G.~Ross and J.~Santiago,
Nucl.\ Phys.\ B {\bf 584} (2000) 313.

\bibitem{LW}
L Cotta-Ramusino, MPhil thesis "Low energy effective theory for brane 
world models" (University of Portsmouth, 2004) 
 
\bibitem{SMS}
T.~Shiromizu, K.~Maeda and M.~Sasaki, Phys. Rev. {\bf D62}, (2000) 024012.

\bibitem{Many}
H.~Hatanaka, M.~Sakamoto, M.~Tachibana and K.~Takenaga,
Prog.\ Theor.\ Phys.\  {\bf 102} (1999) 1213.

\bibitem{antiD1}
S. Mukohyama, hep-th/0505042.

\end{references}
\end{document}